\documentclass[preprint,preprintnumbers,amsmath,amssymb]{revtex4}
\usepackage{graphicx}
\usepackage{dcolumn}
\usepackage{bm}

\begin{document}

\title{The Relativistic framework of Positioning systems}

\author{ J.-F. Pascual-S\'anchez}
\affiliation{Dept. Matem\'atica Aplicada, Facultad de Ciencias,
 Universidad de Valladolid, Valladolid, 47005, Spain, EU}

 \email{e-mail: jfpascua@maf.uva.es}

\begin{abstract}

Emission relativistic coordinates are a class of spacetime coordinates defined
and generated by four emitters (satellites, pulsars) broadcasting their proper
 time by  radio signals. They are the main ingredient of the
 simplest conceivable relativistic positioning system. The emission coordinates
 are independent of any observer.
  Receiving directly the proper time at emission of four satellites, any user
   or observer can measure the values of the emission coordinates, from which
   he/she can obtain  his trajectory and hence, in particular, his position.
   Moreover, if and only if the  four satellites also broadcast to the users the proper times they
are receiving by cross-link autonavigation from the other emitters,
the positioning system is called autolocated or autonomous. In an autolocated positioning
system   the trajectories of the
satellites of the constellation  can also be known by the users and they can also obtain
 the metric of the spacetime (the gravitational field)  on the constellation.

The study of autolocated relativistic positioning systems has been initiated by
Coll and collaborators several years ago and it has been aimed for
developing an exact fully relativistic theory of positioning systems
and gravimetry, based on the framework and concepts of General
Relativity. This exact relativistic framework is the alternative to
 considering post-newtonian relativistic corrections in a classical
 Newtonian framework, which is the customary approach yet now used in
 the   GPS and the   GLONASS.

\end{abstract}
\maketitle

\section{INTRODUCTION}
Practically all observations and experiments in General Relativity are interpreted
 in a
classical Newtonian conceptual framework. In this talk I will focus
on the essential differences between a Newtonian plus post-newtonian
relativistic corrections framework, as that of the current Global
 Navigation
Satellite Systems (GNSS), and a fully relativistic framework which
would be desirable to implement in the future Galileo system due to
its theoretical and practical advantages. The Newtonian conceptual
framework uses  a 3-dim spatial reference system and a time
reference, separately, as still appears  in the   Resolution 3 of
the XXVI General Assembly  of the International
 Astronomical Union, I.A.U., held in Prague last year \cite{iau}. In this framework,
  the "relativistic effects"
are added with the same status as any non-desired perturbation
 (gravitational
influence of other planets, effects of the atmosphere,...). This is made with
 corrections, coming from general relativity when
compared with newtonian gravity, in weak gravitational fields and
with small velocities (post-newtonian formalism).

Typically, this
is what is done nowadays in all the current GNSS, where the primary relativistic
 effects, i.e., some post-newtonian corrections of
second order $1/c^2$, coming from both Special and General Relativity must be taken
 into account. In general, the satellites of the
GNSS are affected by Relativity in three different ways: in the
equations of motion,
in the signal propagation and in the beat rate
of the satellite clocks.

 In the first part of the talk I will review the main clock effects
  at second order, because they are the only
measurable ones in the current GNSS due to the  accuracy of
nanoseconds of the satellite clocks of the GPS and GLONASS systems.
At present, the above approach is perfectly justified from a
practical and numerical point of view and
 it has been successful applied to the GPS, see \cite{ash}.
 However, if the time resolutions are increased (more
accurate clocks), it will be necessary in the future to consider other effects
 of order $1/c^2$ and also it will be necessary to consider
other effects of third (as the Shapiro delay) or fourth order (as the Lense-Thirring
 effect), see \cite{LIN}.

 In this situation, it can be wondered if it
would not be more convenient to change the present newtonian
framework to an exact formulation in full General Relativity. This
would imply to abandon the classical post-newtonian framework.

However, a project for the Galileo system is aimed to develop a theory of positioning
systems in the framework of general relativity. This project, which
is still in a state of theoretical construction, is called SYPOR (a
French acronym for SYstéme de POsitionnement Relativiste). This
project is based in the following mathematical result obtained by
Coll and Morales some time ago \cite{coll1}  which is almost
unknown by the navigation and timing scientific communities, in spite
of its fundamental
 importance in the construction of positioning
systems: {\it In the 4-dim Newtonian spacetime there exists 4, and
only 4, causal classes
 of reference frames, whereas in the relativistic
4-dim Lorentzian spacetime, due to the freedom introduced by the finite propagation
 of light, there exists 199, and only 199, causal
classes of reference frames}.

In general, a causal class is defined by a spacetime frame, a dual coframe and
the 2-dim surfaces generated by the vectors of the frame. Only a
causal class, among the 199 Lorentzian ones,
 is privileged to construct a generic (valid for a wide class
of spacetimes), gravity free (the previous knowledge of the gravitational
field is not necessary) and immediate (non retarded)
positioning system, this is the causal class of the emission coordinates.

As a natural frame is  the set of derivations along the
parameterized lines of a coordinate system, the definition of a determinate causal
class extends  to the coordinate system itself. However,  because the
causal class of a coordinate system is the causal class of its
natural frame at every point and hence local, a coordinate system
may present different causal classes at different points of its
domain of definition and it is said to be inhomogeneous.

The generalized use so far of coordinate systems of the standard Newtonian causal
 class, with  a timelike coordinate parameter  and three spacelike ones,
  exaggerates the partial interest of the newtonian evolution vision 3+1 \`a la ADM, i.e.
 the time evolution  of spatial 3-hypersurfaces, or with the splitting
  1+3 by timelike congruences of observers, in the description of
the coupled spacetime changes of the physical systems.

On the other hand, emission coordinates belong to a causal class of spacetime coordinates
defined and generated by four emitters (satellites) broadcasting their proper
 time by means of radio signals. In other words, the broadcasted signals are
 the emission  coordinates by themselves. These emission
coordinates are covariant (frame independent) and completely
independent of any observer or user. In principle as they define a
positioning system, no observers are necessary at all and hence there
is no necessity of any synchronization procedure. However, any
observer can measure the values of the emission coordinates which
can give his position in the $R^4$ grid of emission coordinates.

Moreover, if and only if the emitters also broadcast to the users the proper times they
are receiving by cross-link autonavigation from the other emitters,
the system is called autolocated and the trajectories of the
emitters can also be known by the users and they can also obtain the metric of the
spacetime (the gravitational field) acting on the constellation.

The study of autolocated positioning systems has
been initiated by Coll and collaborators several years ago
and it has been aimed for developing an exact fully
relativistic theory of positioning systems and gravimetry, based on
the framework and concepts of General Relativity. This is completely different to
 considering post-newtonian ``relativistic effects'' in a classical
 Newtonian framework, which is the customary approach yet now used in the
 present GNSS.

  Just now important results have
been obtained by Coll and collaborators in 2 spacetime dimensions, and some very
 interesting properties have been already
obtained for 3 and 4 spacetime dimensions. I will review them in the second part of the talk.
All the exposition will be mainly based on  \cite{pas}.

\section{The Newtonian framework: relativistic effects in GNSS}
In the GPS system the satellite clocks keep time to an accuracy of
about 4 ${ \rm ns}$ per day. This means once an atomic clock of  the
satellites vehicles (SV)
 is set, then after one day it
should be correct to within about 4 nanoseconds. Therefore, the
errors introduced by the relativistic effects at post-newtonian
order $c^{-2}$ were crucial in the  conception of the current GNSS
and  in its correct operation nowadays because these effects  are of
the order of hundred or thousand nanoseconds.

To describe these errors we could start from a gravitational metric
near the Earth
 that comes from  the
Equivalence Principle (local equivalence of inertia and gravity for
experiments not sensitive to tidal forces), that reads:
\begin{equation}\label{1}
ds^2 = c^2\,d\tau^2=
\left(1-\frac{2GM}{rc^2}\right)\,c^2\,dt^2 -
 dr^2 -r^2 d\psi^2~,
  \end{equation}
and it is not necessary to begin from the full nonlinear
Schwarzschild metric   or even from its linear
approximation. Let us consider the special case of circular
orbits for the satellites of the constellation. Thus the clock on the
Earth and the satellite clock travel at constant distance around
the Earth center, therefore $dr = 0$ for each clock. For both
clocks, one obtains:
\begin{equation}\label{2}
\frac{1}{c^2}\left(\frac{ds}{dt}\right)^2 =
\left(1+\frac{2\,\phi}{ c^2 }\right)\,-\,\frac{v^2}{c^2}\,,
\end{equation}
where $\phi$ is the   Newtonian potential  and
   $v = r \frac{d\psi}{dt}$ is the coordinate
 tangential speed along the circular
equatorial orbit, measured by using the  far-away coordinate time
$t$.

Now, let us apply   (\ref{1}) twice, first
  to the clock in a satellite (S), (using $r = r_{_{S}}$, $v = v_{_{S}}$ and proper
time $d\tau_{_{S}}= ds/c$) and secondly to a fixed clock in the
Earth's equator and turning with it (E) (using, $r = r_{_{E}}$, $v
= v _{_{E}}$ and  proper time $d\tau_{_{E}}=ds/c$), with the same
 elapsed coordinate time, $dt$, corresponding to an inertial observer at spatial
infinity. Then,  dividing both expressions one obtains
at linear order:
\begin{equation}\label{4}
\frac{d\tau_{_E}}{d\tau_{_S}}=
1-\displaystyle\frac{GM}{r_{_{E}}\,c^2}\,-\,\frac{v^2_{_{E}}}{2\,c^2}
+\displaystyle\frac{GM}{r_{_{S}}\,c^2}\,+\,\frac{v^2_{_{S}}}{2\,c^2}\,.
\end{equation}

In (\ref{4}) two main effects appear: 1.-First: {\it Einstein
gravitational shift effect}. The clocks go forward or their time run
faster when they are far away from a center of  gravitational
attraction hence  the time scale is slower for the Earth bound
clock. 2.-Second: {\it Special relativistic Doppler time dilation}.
As the satellite clocks move faster than the clocks on the Earth's
surface this gives rise to the special relativistic Doppler effect
of second order in speed. This effect gives place to
 a red shift in frequency for a signal sent downward from the
satellites to Earth.

Therefore, {\sl this kinematic Doppler effect of second order
 ``works against"  the Einstein gravitational violet shift effect}. To the altitude
  of the satellites of the GPS, $H_{GPS} = \,
 20,183 \, \,{\rm km}$, the Einstein  gravitational shift effect  is
 greater than special relativistic Doppler one.

The main relativistic effects we are considering  are of order
$O(c^{-2})$. This is the order of approximation used in the GPS. So,
why is this order of approximation  good enough? Because, as  said
above, the accuracy of the SVs atomic clocks is nanoseconds and only
these two effects are of higher magnitude.

 Introducing numerical
values in (\ref{4}), the net result for the GPS system is  of the
order of $39,000 \, \, {\rm ns}$ per day when measured in a
laboratory in the rotating Earth. This figure is tantamount to an
error in distance of 11,700 meters after a day of operation. As one
will exposed  in the talk, the previous net result, in which only
local proper times appear, must also be corrected by  using the so-called
global GPS coordinate time and a more realistic gravitational
potential for the Earth. To do this, we must restart with a more
realistic metric in the weak external gravity field of the Earth
which is obtained from the linear Schwarzschild metric including the
quadrupole moment in the gravitational potential, the centrifugal
potential and the kinematical Sagnac effect due to Earth rotation,
\cite{ash, pas}.

On the other hand, it is important to point out that the necessity
to consider  other smaller relativistic effects
 at the order $O(c^{-3})$  when laser cooled atomic clocks
   of last  generation   are
used, with measurable errors of  order of picoseconds, as in the
ACES (Atomic Clock Ensemble in Space) mission of the ESA for the
International Space Station, ISS, and those of the order $O(c^{-4})$
\cite{LIN},  when one
 works  with intrinsic errors of femtoseconds in the near future.

\section{The Relativistic framework of Positioning systems}

\subsection{Reference systems and positioning systems}
Let us define location systems as the physical realizations of some
coordinate systems. Location systems are of two different types:
reference systems and positioning systems. The first ones are
4-dimensional reference systems which allow one observer, considered
at the origin, to assign four coordinates to the events of his
neighborhood  by means of  a radio signal. Due to the finite speed
of light, this assignment is retarded with a time delay. The second
ones are 4-dimensional positioning systems (as  intended to be used
in the SYPOR project) which allow to every event of a given domain
to know its proper coordinates  in an immediate way.

In Relativity, a (retarded) reference system can be constructed
starting from an (immediate) positioning system (it is sufficient
that each event sends its coordinates to the observer at the origin)
but not the other way around. In contrast, in Newtonian theory,
3-dimensional reference and positioning systems are interchangeable
and as the velocity of transmission of information, the speed of
light, is supposed to be infinite, the Newtonian reference systems
are not retarded but immediate.

The reference and positioning systems defined here are 4-dimensional
objects, including time location.

Following \cite{coll1}, the best  way to visualize and characterize
a space–time coordinate system is to start from four families of
coordinate 3-surfaces, then, their mutual intersections give six
families of coordinate 2-surfaces and four congruences of coordinate
lines. Alternatively,  one can use the related covectors or 1-forms
$ \{ {\bm \theta}^i\}, i = 1,2,3,4$, instead of the 3-surfaces, and
the vectors of a coordinate tangent frame $ \{{\bm e}_i\}, i =
1,2,3,4$,  instead of four congruences of coordinate lines which are
their integral curves. In this way, for a specific domain of a
Lorentzian or Newtonian spacetime, each coordinate system is fully
characterized by its causal class, which is defined by a set of 14
characters:
\begin{equation}\label{22}
\{{\rm{c_1\, c_2\, c_3\, c_4, C_{12}\, C_{13}\, C_{14}\, C_{23}\,
C_{24}\, C_{34}}},{\it c_1\, c_2\, c_3\, c_4}\},
\end{equation}
being ${\rm{c_i}}$ the Lorentzian causal character of the vector
${\bm{e}_i}$, i.e. if it is spacelike, timelike or lightlike;
${\rm{C_{ij}}}$ the causal character of the adjoint 2-plane
$\{{\bm{e_}i\wedge\,\bm{e_}j\}}$ and, finally, $\it{c_i}$ the causal
character of the covectors ${\bm \theta}^i$ of the dual coframe,
 ${\bm \theta}^i({\bm{e}_j})=\delta^i_j$. The covector ${\bm \theta}^i$  is
 timelike (resp. spacelike) iff the 3-plane generated by the
 three vectors $\{{\bm{e}_j}\}_{j\neq i}$ is spacelike (resp.
 timelike). This applies for both Newtonian and Lorentzian
 spacetimes. In addition, for the latter, the covector ${\bm \theta}^i$
 is lightlike iff the 3-plane generated by $\{{\bm{e}_j}\}_{j\neq
 i}$ is lightlike or null.

 This new degree of freedom (lightlike) in the causal
 character, which is proper of Lorentzian relativistic spacetimes
 but which does not exist in Newtonian spacetimes,  allows to obtain in \cite{coll1}, as it has been commented
  in the introduction, the following conclusion: {\sl In the 4-dimensional Newtonian
   spacetime there exist four, and only four, causal classes of frames, whereas in the
    relativistic 4-dimensional Lorentzian spacetime,
   due to the freedom commented above, there exists 199,
   and only 199, causal classes of frames.}

As notation for the causal characters, we will use lower case roman
types ({\rm{s,t,l}}) to represent the causal character of vectors
(resp. spacelike, timelike, lightlike), and capital types (S,T,L)
and lower case italic types ({\it s,t,l}) to denote the causal
character of 2-planes and covectors, respectively.

In Relativity, a specific causal class, among the 199 ones, can be
assigned to any of the different  coordinate systems used in all the
solutions of the Einstein equations. However, for the same
coordinate system and the same solution, the causal class can change
depending on the region of the spacetime considered and the coordinate system is said inhomogeneous.

 A coordinate system can be constructed starting from a tangent
frame or its coordinate lines. But  in this way, in general, there
are obstructions to obtain   positioning systems that are generic,
i.e., valid for many different spacetimes. On the other hand,
generic positioning systems can be constructed  from four
one-parameter families of 3-surfaces or, equivalently, from a
cotangent frame. One clock broadcasting its proper time is described
in the spacetime by a timelike line in which each event is the
vertex of a future light cone. The set of these light cones of a
emitter constitutes a one-parameter (proper time) family of null
hypersurfaces. So, four clocks broadcasting their proper times
determine four one-parameter families of null or lightlike
3-surfaces.

In a relativistic spacetime, the wave fronts of those signals
parameterized by the proper time of the clocks, define four families
of light cones (3-dimensional null hypersurfaces) which contain all
the lightlike geodesics of four emitters and  making a contravariant
null (or light) coordinate system \cite{coll1}, see Fig. 1. Such a
coordinate system does not exist in a Newtonian spacetime where the
light travels at infinite speed.

\subsection{Coll positioning systems}
 Among the 199 Lorentzian causal classes, only one is privileged
to construct a generic (valid for a wide class of spacetimes),
gravity free (the previous knowledge of the gravitational field is
not necessary) and immediate positioning system.  This is the causal
class $\{{\rm{s\,s\,s\,s,S\,S\,S\,S\,S\,S}}, l\,l\,l\,l \}$ of the
Coll-Morales homogeneous coordinate system \cite{coll1,pas,coll2}. In this class
the {\sl emission coordinates} of the Coll positioning systems are
included. These emission coordinates have been also studied in
\cite{rovelli,hehl} in the special case of a flat Minkowski
spacetime without gravity.

The coordinate system of this causal class is always homogeneous and it has associated four
families of null 3-surfaces or a real null coframe, whose mutual
intersections give six families of spacelike 2-surfaces and four
congruences of spacelike lines.

   In this primary positioning
system, an user at any event in a given spacetime region can know
its proper coordinates. The four proper times of four satellites
 $ (\{\tau^A \};\, A = 1,2,3,4$) read at an event by a receiver or user
constitute the null (or light) proper emission coordinates or user
positioning data of this event, with respect to four SVs, see Fig.
2. These four numbers can be understood as the ``distances" between
the event and the four satellites.

In  a certain domain
$\Omega\subset\mathbb{R}^4$ of the grid of parameters $ (\{\tau^A
\};\, A = 1,2,3,4$), any user receiving continuously his emission
coordinates may know his trajectory. If the observer has his own
clock, with proper time denoted by $\sigma$, then he can know his
trajectory, $\tau^A=\tau^A(\sigma)$, and  his four-velocity,
$u^A(\sigma)= d \tau^A/d\sigma$.
\begin{figure}
\begin{center}
 \includegraphics[width=3in]{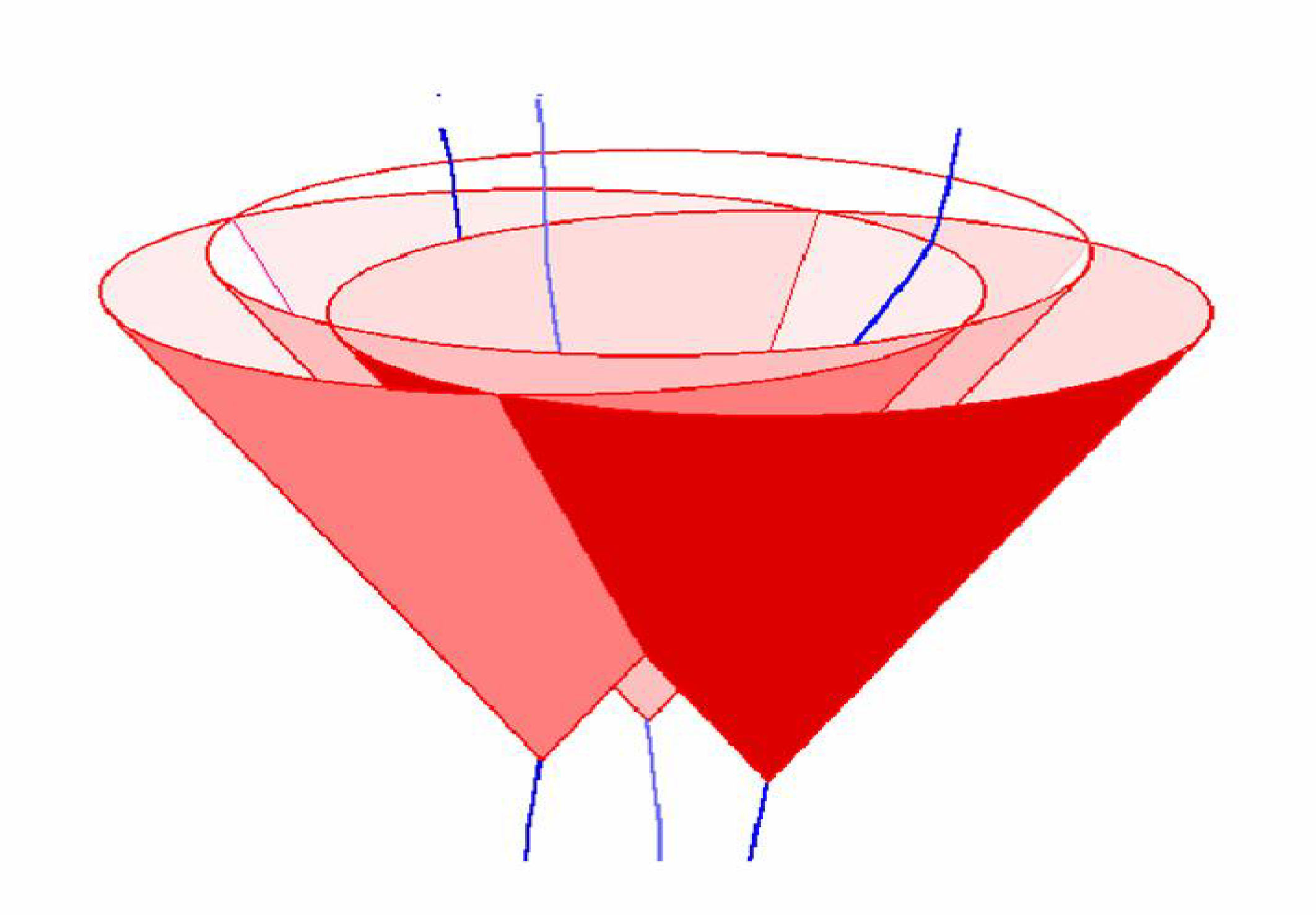}
\end{center}
\caption{{\small  Proper relativistic emission coordinates:
intersection of the four future light cones of the SVs with  the
past light cone of a receiver. In the Figure only 3 light cones of
the SVs are drawn in a Lorentzian spacetime of 1+2 dimensions.}}
\end{figure}
There is not space-time asymmetry like in
the standard Newtonian coframe  $(\it{t\,s\,s\,s })$ (one
timelike ``$\it{t}$" and three  spacelike ``$\it{s}$''). In
emission coordinates  obtained from a general real null coframe
  $ (\it{l\,l\,l\,l})=\{d\tau^1, d\tau^2, d\tau^3, d\tau^4\}$, which is neither orthogonal nor
   normalized, the contravariant spacetime metric is symmetric with null diagonal
  elements  and it has the general expression \cite{coll5}:
\begin{equation}\label{25}
(g^{AB})=  ({d\tau ^A} \cdot d\tau ^B) =\left(\begin{array}{cccc} 0 &g^{12}& g^{13}&g^{14}\\
g^{12}&0& g^{23}&g^{24}\\
g^{13}& g^{23}&0&g^{34}\\
g^{14}& g^{24}&g^{34}&0\end{array}\right),
\end{equation}
where $g^{AB}>0$ for $A\neq B$. Four null covectors can be
linearly dependent although none of them is proportional to
another. To ensure that the four null covectors are linearly
independent and span a 4-dim spacetime, it is sufficient that
$\text{det}(g^{AB})\neq0$. Finally, this metric has a Lorentzian
signature $(+,-,-,-)$ iff $\text{det}(g^{AB})<0$. The expression
(\ref{25}) of the metric is observer independent and has six
degrees of freedom.

A splitting of this metric can be considered,
changing from the six independent components (ten components minus four gauge degrees
of freedom of passive coordinate transformations) of $g^{AB}$  to a more
convenient set, which neatly separates two shape parameters depending
only on the direction of the covectors $d\tau^A$ or equivalently depending exclusively on the
 trajectories of the emitters, from other four scaling
parameters depending on the length of the covectors or depending on the proper time
of each satellite.
\begin{figure}
\begin{center}
  \includegraphics[width=2.5in]{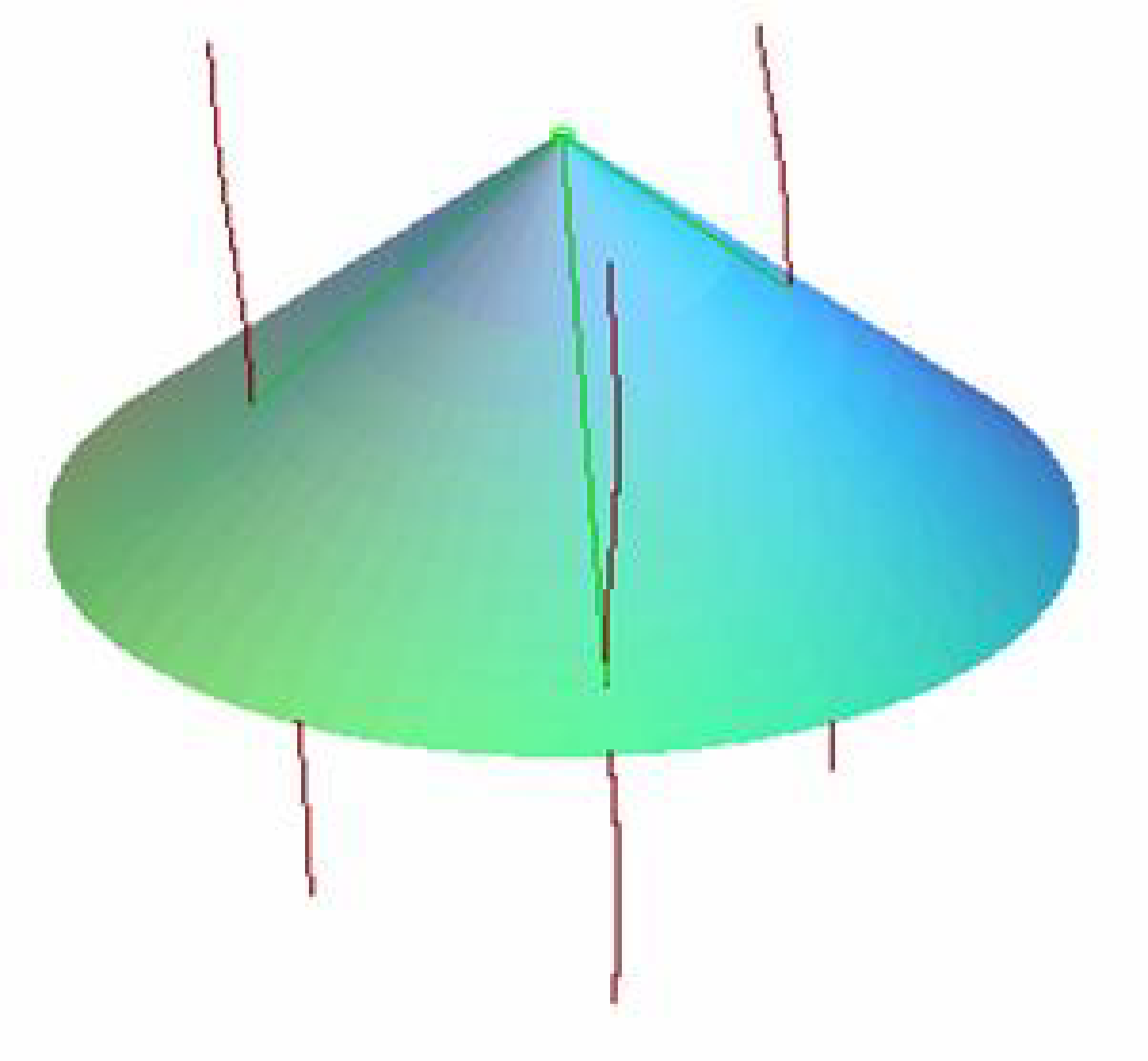}
\end{center}
\caption{{\small  Past light cone of an event in  1+2 dimensions,
the proper time parameterized paths of 3 SVs (in red) and the
lightlike geodesics (in green) followed by the signal from each
satellite to a event of the trajectory of a receiver.}}
\end{figure}
Four satellites emitting, without the necessity of a
synchronization convention,  not only their proper times $ \tau
^A$, but also the proper times $ \tau^{AB}$ of  three close
satellites received by the satellite $A$ in $\tau^A$ (in total $
\{\tau^A, \tau^{AB}\};\, A \neq B; \,A, B = 1,2,3,4$) and transmitted, constitute
an autonomous or autolocated positioning system. This is because  the three
proper times, together with  its proper time, that a satellite
clock receives from other SVs constitute its proper emission
coordinates.

The sixteen data $ \{\tau^A, \tau^{AB} \} $ or emitter positioning
data received by a user, allows the user to know his/her
trajectory, the trajectories of four SVs and  the metric of the spacetime
 (the gravitational field) acting on the constellation.

Coll positioning systems  are yet now quite well developed
for two-dimensional spacetimes \cite{coll3, coll4} for arbitrary and special observers
in Minkowski and Schwarzschild spacetimes.
 However, the known results
for the two dimensional case
are not trivially generalizable for the realistic four dimensional one \cite{coll5,rey}
and much work remains to be done.

Finally it is to be pointed out that, although the main idea of positioning systems based
 on emission coordinates
 is simple, our partial lack of intuition at
present about the general use of  emission coordinates is usually accompanied
 by many  prejudices coming from the newtonian framework,  basically from the
  newtonian evolution vision 3+1 \`a la ADM.

\section*{Acknowledgments}
  I am grateful to Bartolom\'e  Coll by communication along the latter years.
  Also I thank to Joan Josep Ferrando and Juan Antonio Morales by some friendly
 and humorous discussions.
 Finally, I am thankful to Angel San Miguel and Francisco Vicente by collaboration.  The author is
currently partially supported by  the Spanish Ministerio de
Educaci\'on y Ciencia MEC ESP2006-01263 with EU-FEDER funding and
Junta de Castilla y Le\'on VA065A05 grants.

\end{document}